\documentclass[twoside]{dis07}
\usepackage[latin1]{inputenc}
\usepackage[dvips]{graphicx,epsfig,color}
\usepackage{psfrag,wrapfig,rotating}
\usepackage{amssymb,amsmath,array}

\usepackage{epsf}
\usepackage{amsfonts}
\usepackage{cite}
\usepackage[small]{caption2}
\usepackage{graphics}

\def\slashchar#1{\setbox0=\hbox{$#1$}
   \dimen0=\wd0
   \setbox1=\hbox{/} \dimen1=\wd1
   \ifdim\dimen0>\dimen1
      \rlap{\hbox to \dimen0{\hfil/\hfil}}
      #1
   \else
      \rlap{\hbox to \dimen1{\hfil$#1$\hfil}}
      /
   \fi}

\newcommand{\be}{\begin{equation}}
\newcommand{\ee}{\end{equation}}
\newcommand{\eq}{\end{equation}}
\newcommand{\rb}{\underline{r}}
\newcommand{\kb}{\underline{k}}

\newcommand{\scalingm}{1.3cm}

\newcommand{\scaling}{2.1cm}

\newcommand{\scalingA}{1.5cm}
\newcommand{\scalingB}{2.2cm}
\newcommand{\scalingC}{2.6cm}
\newcommand{\scalingD}{2.2cm}

\begin{document}
\title{QCD factorizations in $\gamma^{*} \gamma^{ *} \to \rho_L^0 \rho_L^0$}

\author{M.SEGOND
%
%
\vspace{.3cm}\\
%
LPT \\
-Universit\'e Paris-Sud-CNRS, 91405-Orsay,  France
}
%


\maketitle

\begin{abstract}
The exclusive reaction of rho meson pair electroproduction in $\gamma^*\gamma^*$ collisions is a nice place to study various dynamics and factorization properties in the perturbative sector of QCD. At low energy (quarks dominance), this process can be considered as a way to explore QCD factorizations involving generalized distribution amplitudes (GDA) and transition distribution amplitudes (TDA), and, in the Regge limit of QCD (gluons dominance), it seems to offer a promising probe of the BFKL resummation effects which could be studied at the next international linear collider (ILC).
\end{abstract}

\section{GDA/TDA factorizations at low energy}
\label{fact1}

\subsection{The Born order amplitude}
\label{secBorn}

We calculate \cite{gdatda} the scattering amplitude of the process $ \gamma^*(q_1)
\gamma^*(q_2) \to \rho^0_L (k_{1})\rho^0_L(k_{2})$  at Born order for both transverse and longitudinal polarizations  in the forward kinematics, when quark exchanges dominate. The virtualities $Q^2_i=-q^2_i ,$ supply the hard scale which justifies the perturbative computation of the amplitude $M_H$. The final states $\rho$ mesons are described in the collinear factorization by  their distribution amplitudes (DA) in a similar way as in the classical work of Brodsky-Lepage \cite{BLphysrev24}.

\subsection{$\gamma_T^* \gamma_T^* \to \rho^0_L \rho^0_L$ in the generalized Bjorken limit}
\label{secgda}

\begin{figure}[htb]
\psfrag{r1}[cc][cc]{$\quad\rho(k_1)$}
\psfrag{r2}[cc][cc]{$\quad\rho(k_2)$}
\psfrag{p1}[cc][cc]{}
\psfrag{p2}[cc][cc]{}
\psfrag{p}[cc][cc]{}
\psfrag{n}[cc][cc]{}
\psfrag{q1}[cc][cc]{$q_1$}
\psfrag{q2}[cc][cc]{$q_2$}
\psfrag{GDA}[cc][cc]{$GDA_H$}
\psfrag{Da}[cc][cc]{DA}
\psfrag{HDA}[cc][cc]{$M_H$}
\psfrag{M}[cc][cc]{$M$}
\psfrag{Th}[cc][cc]{$T_H$}
\centerline{\scalebox{1}
{$\begin{array}{cccccc}
\!\!\raisebox{-0.44\totalheight}{\epsfig{file=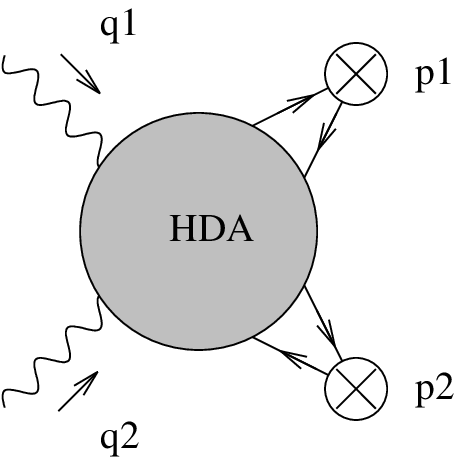,width=\scalingD}}& \begin{array}{c}
\raisebox{0.85 \totalheight}
{\psfrag{r}[cc][cc]{$\; \; \quad\rho(k_1)$}
\psfrag{pf}[cc][cc]{}
\hspace{-.4cm}\epsfig{file=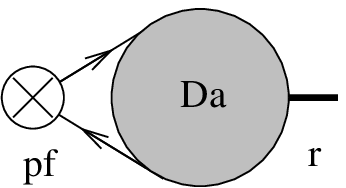,width=\scalingA}}\\
\raisebox{-0.65 \totalheight}
{\psfrag{r}[cc][cc]{$\; \; \quad\rho(k_2)$}
\psfrag{pf}[cc][cc]{}
\hspace{-.4cm}\epsfig{file=segond_mathieu.fig1_DA.eps,width=\scalingA}}
\end{array}&=& \!
\raisebox{-0.44 \totalheight}
{\psfrag{r}[cc][cc]{$\; \; \quad\rho(k_1)$}
\psfrag{pf}[cc][cc]{}
\epsfig{file=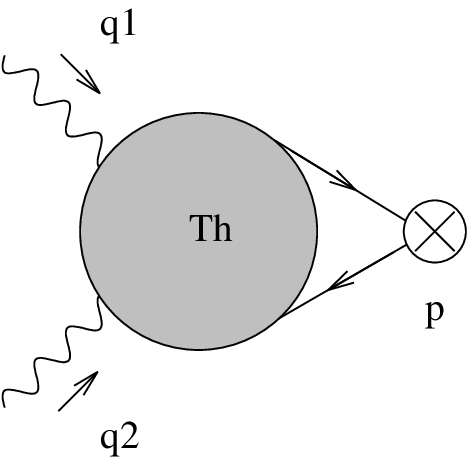,width=\scalingB}}&
\raisebox{-0.43\totalheight}{\epsfig{file=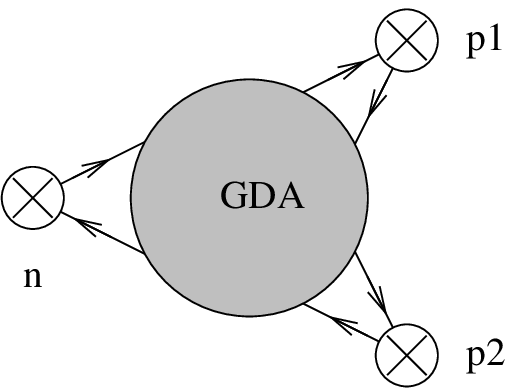,width=\scalingC}}& \begin{array}{c}
\raisebox{0.99 \totalheight}
{\psfrag{r}[cc][cc]{$\; \; \quad\rho(k_1)$}
\psfrag{pf}[cc][cc]{}
\hspace{-.4cm}\epsfig{file=segond_mathieu.fig1_DA.eps,width=\scalingA}}\\
\raisebox{-0.4 \totalheight}
{\psfrag{r}[cc][cc]{$\; \; \quad\rho(k_2)$}
\psfrag{pf}[cc][cc]{}
\hspace{-.4cm}\epsfig{file=segond_mathieu.fig1_DA.eps,width=\scalingA}}
\end{array}
\end{array}
$}}

\caption{Factorization of the  amplitude in terms of a GDA which is expressed in a perturbatively computed $GDA_H$ convoluted with the DAs of the two $\rho$-mesons. }
\label{FactGDA}
\end{figure}

We then consider transverse photons whose  scattering energy is much smaller than the typical scales of the process (close to the semi-exclusive limit in DIS when $x_{Bj}\to1$). We obtain the same expression of the amplitude computed previously (Sec.~\ref{secBorn}) in a different theoretical framework which is based on the factorization property of the scattering amplitude in terms of a hard coefficient function $T_H$ convoluted with a GDA encoding the softer part of the process, as illustrated in Fig.~\ref{FactGDA}.

 \vspace{.8cm}
\begin{wrapfigure}{r}{0.35\columnwidth}
\psfrag{q1}[cc][cc]{$q_1$}
\psfrag{q2}[cc][cc]{$q_2$}
\psfrag{p1}[cc][cc]{}
\psfrag{p2}[cc][cc]{}
\psfrag{n}[cc][cc]{}
\psfrag{p}[cc][cc]{}
\psfrag{Tda}[cc][cc]{$TDA_H$}
\psfrag{Th}[cc][cc]{$T_H$}
\psfrag{Da}[cc][cc]{DA}
\centerline{\scalebox{1.05}
{
$\begin{array}{c}
\begin{array}{cc}
\raisebox{-0.44 \totalheight}{ \hspace{-.6cm}\epsfig{file=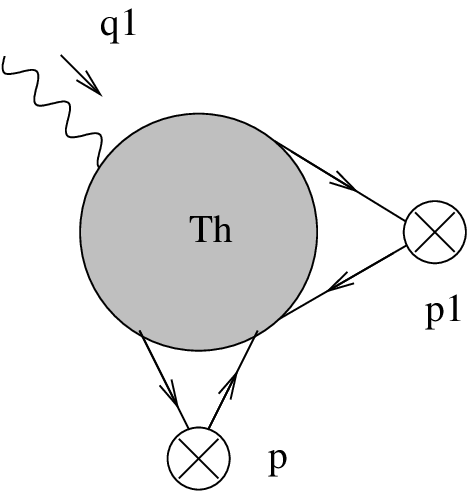,width=\scaling}}&
\raisebox{-0.2\totalheight}
{\psfrag{r}[cc][cc]{$\qquad \rho(k_1)$}
\psfrag{pf}[cc][cc]{ \raisebox{-1.05 \totalheight}{}}\hspace{-.2cm}\epsfig{ file=segond_mathieu.fig1_DA.eps,width=\scalingm}}
\end{array}\\
\\
\begin{array}{cc}
\raisebox{-0.34 \totalheight}{\hspace{-.6cm}\epsfig{file=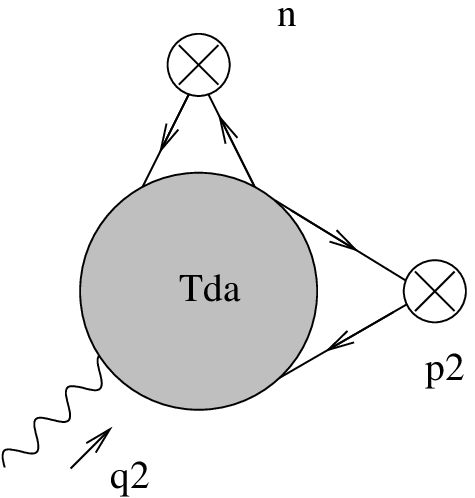,width=\scaling}}&
\raisebox{-0.24\totalheight}
{\psfrag{r}[cc][cc]{$\qquad \rho(k_2)$}
\psfrag{pf}[cc][cc]{\raisebox{-1.05 \totalheight}{}}\hspace{-.2cm}\epsfig{file=segond_mathieu.fig1_DA.eps,width=\scalingm}}
\end{array}
\end{array}
$}}
\caption{Factorization involving a TDA  which is written  as the convolution of a hard term $TDA_H$ and a DA of the $\rho$-meson.\vspace{-1.05cm}}
\label{FactTDA}
\end{wrapfigure}

 \vspace{-0.9cm}
\subsection{$\gamma_L^* \gamma_L^* \to \rho^0_L \rho^0_L$
with strong ordering of virtualities }
\label{sectda}
In the regime with strong ordering of the virtualities $Q_{1}^2  \gg  Q_{2}^2  $, we compute the amplitude with initial longitudinally polarized  photons, in a factorized formula involving a convolution of a hard coefficient function $T_H$ and
a $\gamma^* \to \rho$  TDA. This soft part is defined with the leading twist quark-antiquark non local correlator between non-diagonal matrix elements corresponding to the $\gamma \to \rho$ transition.  We also obtain the same expression as in the direct calculation of the Sec.~\ref{secBorn} in this kinematics.


\section{$k_\perp$-factorization in the Regge limit of QCD}

\subsection{Impact factor representation}
\label{sec:figures}

\psfrag{p1}[cc][cc]{$k_1$}
\psfrag{p2}[cc][cc]{$k_2$}
\psfrag{q1}[cc][cc]{$q_1$}
\psfrag{q2}[cc][cc]{$q_2$}
\psfrag{l1}[cc][cc]{}
\psfrag{l1p}[cc][cc]{}
\psfrag{l2}[cc][cc]{}
\psfrag{l2p}[cc][cc]{}
\psfrag{r}[cc][cc]{$r$}
\begin{wrapfigure}{r}{0.35\columnwidth}
\centerline{\includegraphics[width=0.3\columnwidth]{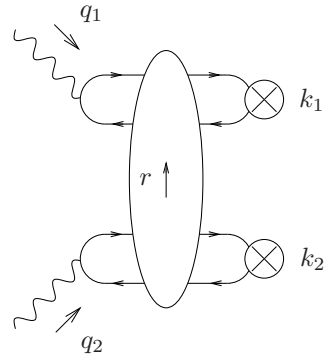}}
\caption{\small The amplitude of the process $\gamma^*_{L,T}(q_1) \gamma^*_{L,T}(q_2) \to \rho_L^0(k_1)  \rho_L^0(k_2)$
in the impact representation.\vspace{-0.85cm}}
\label{impact}
\end{wrapfigure}
We are focusing now on the high-energy (Regge) limit, when the cm energy $s_{\gamma^*\gamma^*}$ is much larger than all other scales of the process, in which $t-$channel gluonic
exchanges dominate \cite{born}. 
The  highly virtual photons provides ones small transverse size objects ($q \bar{q}$ color dipoles) whose scattering is the cleanest place to study the typical Regge behaviour with $t-$channel  BFKL Pomeron exchange \cite{bfkl}, in perturbative QCD. 
  If one selects the events with
comparable photon virtualities, the BFKL resummation effects dominate with respect to the
conventional partonic evolution of DGLAP \cite{dglap} type. Several studies of BFKL dynamics have been performed at the level of the total cross-section \cite{bfklinc}. At high energy, the impact factor representation of the scattering amplitude 
 has the form of a convolution in the transverse momentum  $\kb$ space between the  two  impact factors corresponding  to the
transition of
$\gamma^*_{L,T}(q_i)\to \rho^0_L(k_i)$ 
via the $t-$channel exchange of two reggeized gluons (with momenta $\kb$ and $\rb -\kb$).



\subsection{Non-forward cross-section at ILC for $e^+e^- \to e^+e^- \rho_L^0  \;\rho_L^0$ }

Our purpose is now to evaluate at Born order and in the non-forward case the cross-section of the process $e^+e^- \to e^+e^- \rho_L^0  \;\rho_L^0$ in the planned  experimental conditions 
of the International Linear Collider (ILC). We focus on the LDC detector project and we use the potential of the very forward
region accessible through the electromagnetic calorimeter  BeamCal which may be installed
around the beampipe at 3.65 m from the interaction point. This calorimeter allows to detect (high energetic) particles down to 4 mrad. This important technological step was not feasible a few years ago.
At ILC, the foreseen cm energy is $ \sqrt{s}=500$ GeV.
Moreover we  impose that $s_{\gamma^*\gamma^*}  >c \, Q_1 \, Q_2$ (where $c$ is an arbitrary constant). It is required by the Regge kinematics for which the impact representation is valid. 
We choose $Q_i$ to be bigger than 1 GeV since it provides the  hard scale of the process. $Q_{i\, max}$ will be fixed to 4 GeV: indeed the various amplitudes
involved are completely negligible for higher values of virtualities.

\begin{wrapfigure}{r}{0.4\columnwidth}
\centerline{\includegraphics[width=0.4\columnwidth]{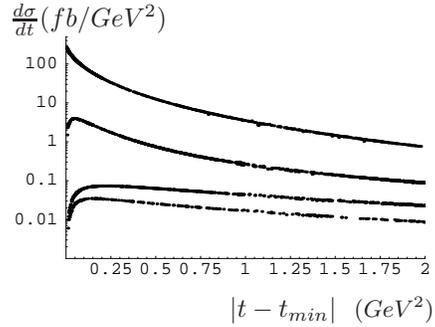}}
\begin{picture}(10,20)
\put(80,10){ $|t-t_{min}| $ \  \small $ (GeV^2)$}
\put(0,120){$\frac{d\sigma}{dt} (fb/GeV^2)$}
\end{picture}
\caption{\small Cross-sections for $e^+e^- \to e^+e^- \rho_L^0  \;\rho_L^0$ process. Starting from above, we display the
 cross-sections  corresponding to the  $\gamma^*_L \gamma^*_L$ mode,  to the $\gamma^*_L\gamma^*_T$ modes,   to the 
$\gamma^*_T \gamma^*_{T'}$ modes with  different $T \neq T'$ and finally to the $\gamma^*_T \gamma^*_{T'}$ modes with  the same $T=T'$.}
\label{FigLogcurves}
\end{wrapfigure}
We now display in Fig.\ref{FigLogcurves} the cross-sections  as a function of the momentum transfer $t$ for the different $\gamma^{*}$ polarizations. For that we performed analytically the integrations over  $\kb$ (using conformal transformations to reduce the number of massless propagators) and numericaly the integration over the accessible phase space.
 We assume the QCD coupling constant to be $\alpha_{s}(\sqrt{Q_1 Q_2})$ running at three loops, the parameter $c=1$ which enters in the Regge limit condition and the 
 energy of the beam $ \sqrt{s}=500$ GeV. We see that all the differential cross-sections which involve at least one transverse
photon vanish in the forward case when $t=t_{min}$, due to the $s$-channel helicity conservation.
We finally display in the Table.\ref{tab} the results for the total cross-section integrated over $t$ for various values of c.  With the foreseen nominal integrated luminosity of $125  \, {\rm fb}^{-1},$ this will yield $4.26\, 10^3$ events per year with $c=1$.

By looking into the upper 
curve in the Fig.\ref{FigLogcurves} related to the longitudinal polarizations, one sees that the point  $t=t_{min}$ gives the maximum of the total cross-section (since the transverse polarization case vanishes at $t_{min}$) and then practically dictates the trend of the total cross-section which is strongly peaked in the forward direction (for the longitudinal case) and strongly decreases with $t$ (for all polarizations).
From now we only consider the forward dynamics.
\begin{wraptable}{l}{0.2\columnwidth}
\centerline{\begin{tabular}{|l|r|}
\hline
c  & $\sigma^{Total} \, ( fb )$ \\\hline  
1  & 34.1 \,  \,    \\\hline
2         & 29.6  \,  \, \\\hline
10             & 20.3  \,  \,    \\\hline
\hline
\end{tabular}}
\caption{Total cross-section for various c.}
\label{tab}
\end{wraptable}

\begin{wrapfigure}{r}{0.5\columnwidth}
\centerline{\includegraphics[width=0.4\columnwidth]{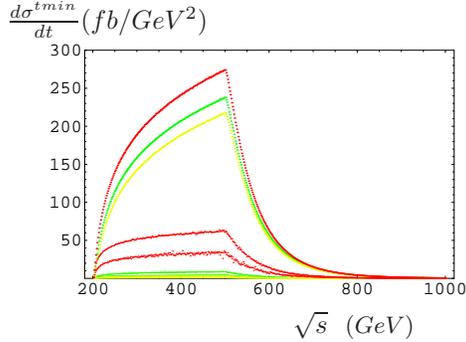}}
\begin{picture}(10,10)
\put(110,0){ $\sqrt{s} $ \  \small $ (GeV)$}
\put(5,115){$\frac{d\sigma^{tmin}}{dt} (fb/GeV^2)$}
\end{picture}
\caption{\small Cross-sections for $e^+e^- \to e^+e^- \rho_L^0  \;\rho_L^0$ at $t=t_{min}$ for different values of the parameter $c$: the red (black) curves correspond to  $c=1$, the green (dark grey) curves to $c=2$ and and the yellow (light grey) curves to $c=3$. For each value of c, by decreasing order the curves correspond to gluon-exchange, quark-exchange with longitudinal virtual photons and quark-exchange with transverse virtual photons. }
\label{Figceffects}
\end{wrapfigure}

The Fig.\ref{Figceffects} shows the cross-section (for both gluons and quarks exchanges) at $t_{min}$ for different values of the parameter c which enters in the Regge limit condition
: the increase of c leads to the suppression of quarks exchanges (studied in section \ref {fact1}) and we base the value of $c$ chosen previously on the gluon exchange dominance over the quark exchange contribution.


The ILC collider is expected to run at a cm nominal energy of 500 GeV, though it might be extended in order to cover a range between 200 GeV and 1 TeV.  Although the Born order cross-sections do not depend on $s,$ the triggering effects introduce an $s$-dependence; note that the cross-section falls down between 500 GeV and 1 TeV. The measurability is then optimal  when $\sqrt{s} = 500$ GeV.
The results obtained at Born approximation can be considered as a lower limit of the cross-sections for $\rho$-mesons pairs production with complete BFKL evolution taken into account. We consider below only the point $t=t_{min}$ and we restrict ourselves to the leading order (LO) BFKL evolution in the saddle point approximation.  

\begin{wrapfigure}{r}{0.45\columnwidth}
 \vspace{-8cm}
\centerline{\includegraphics[width=0.4\columnwidth]{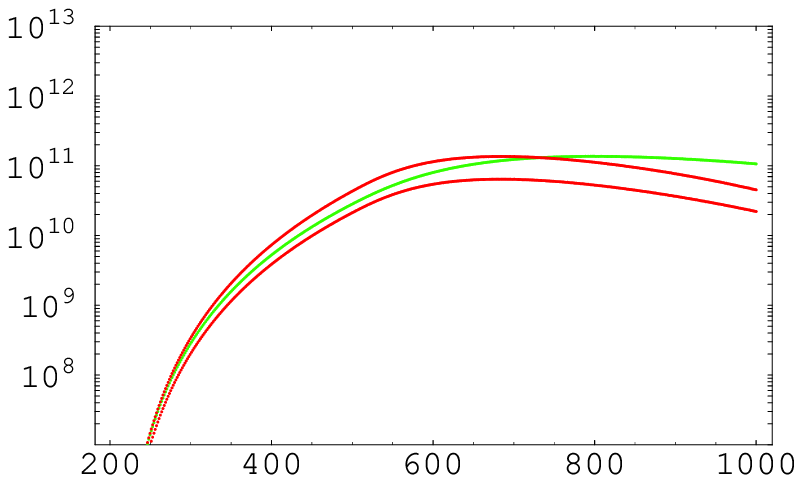}}
\begin{picture}(10,10)
\put(110,0){ $\sqrt{s} $ \  \small $ (GeV)$}
\put(5,115){$\frac{d\sigma^{tmin}}{dt} (fb/GeV^2)$}
\end{picture}
\caption{\small Cross-sections for $e^+e^- \to e^+e^- \rho_L^0  \;\rho_L^0$ with LO BFKL evolution at $t=t_{min}$ for different $\alpha_s$ : the upper and lower  red (black) curves for  $\alpha_s$ running respectively at one and three loops  and the green one for $\alpha_s = 0.46$.}
\label{Figbfkltmin}
\end{wrapfigure}

 From previous studies at the level of $\gamma^*\gamma^*$ \cite{epsw},  the NLO contribution is expected to be between the LO and Born order cross-sections. This ordering will be preserved at the level of the $e^+e^-$ process.
The comparison of Figs.\ref{Figceffects} with  Figs.\ref{Figbfkltmin} leads to the conclusions that the BFKL evolution changes the shape of the cross-section:
 when increasing $\sqrt{s}$
 from 500 GeV to 1 TeV, the two gluon exchange cross-section will fall down, while the cross-section with  the BFKL
resummation effects taken into account should more or less stay stable, with a high number of events to be still observed for these cm energies.

\vspace{0.161cm}

\noindent


\section{Acknowledgments}
\vspace{-.15cm}
This work was done in a collaboration with B.Pire, L.Szymanowski and S.Wallon. I am very gratefull to the DIS 2007 organizers.
\vspace{-.15cm}
\section{Bibliography}
\begin{footnotesize}
\vspace{-.35cm}
Slides: \\ 
\verb$http://indico.cern.ch/contributionDisplay.py?contribId=95&sessionId=7&confId=9499$
 




%

\end{footnotesize}


\end{document}